\documentclass[iop,numberedappendix] {emulateapj}

\usepackage{amsmath}
\usepackage{mathrsfs}
\usepackage{empheq}
\usepackage{url}
\usepackage{multirow}
\usepackage{dcolumn}   
\usepackage{amssymb}   
\usepackage{hyperref}
\usepackage[applemac]{inputenc}
\usepackage{graphicx,subfigure}
\usepackage{bm}
\usepackage{color}
\usepackage[export]{adjustbox}

\shorttitle{Transition from turbulent to dead zone in PPDs}
\shortauthors{Fulvia Pucci, Kengo Tomida, James Stone, Shinsuke Takasao, Shoichi Okamura}

\begin{document}

\title{Transition region from turbulent to dead zone in protoplanetary disks: local shearing box simulations}

\author{Fulvia Pucci\altaffilmark{1,2},  Kengo Tomida\altaffilmark{4}, James Stone\altaffilmark{3}, Shinsuke Takasao\altaffilmark{5}, Hantao Ji\altaffilmark{2}, Shoichi Okamura\altaffilmark{1,6}}
\email{pucci@nins.jp}
\email{fpucci@princeton.edu}
\email{tomida@astr.tohoku.ac.jp}
\email{jmstone@ias.edu}
\email{takasao@astro-osaka.jp}
\email{hji@pppl.gov}
\email{okamura@nifs.ac.jp}
\affil{\altaffilmark{1}International Research Collaboration Center, National Institutes of Natural Sciences, Tokyo 105-0001, Japan}
\affil{\altaffilmark{2} Princeton University, Astrophysics department, US}
\affil{\altaffilmark{3}The Institute for Advanced Study, 1 Einstein Drive, Princeton, NJ 08540, USA}
\affil{\altaffilmark{4} Astronomical Institute, Tohoku University, Sendai, Miyagi 980-8578, Japan}
\affil{\altaffilmark{5} Department of Earth and Space Science, Osaka University, Toyonaka, Osaka 560-0043, Japan}
\affil{\altaffilmark{6} National Institute for Fusion Science, National Institutes of Natural Sciences,
Toki 509-5292, Japan}
\begin{abstract}
The dynamical evolution of protoplanetary disks is of key interest for building a comprehensive theory of planet formation and to explain the observational properties of these objects. Using the magnetohydrodynamics code Athena++, with an isothermal shearing box setup, we study the boundary between the active and dead zone, where the accretion rate changes and mass can accumulate. We quantify how the turbulence level is affected by the presence of a non uniform ohmic resistivity in the radial - x direction that leads to a region of inhibited turbulence (or dead zone). Comparing the turbulent activityto that of ideal simulations, the turbulence inhibited area shows density fluctuations and magnetic activity at its boundaries, driven by energy injection from the active (ideal) zone boundaries. We find magnetic dissipation to be significantly stronger in the ideal regions, and the turbulence penetration through the boundary of the dead zone is determined by the value of the resistivity itself,  through the ohmic dissipation process, though the thickness of the transition does not play a significant role in changing the dissipation. 
We investigate the 1D spectra along the shearing direction: magnetic spectra appear flat at large scales both in ideal as well as resistive simulations, though a Kolmogorov scaling over more than one decade persists in the dead zone, suggesting the turbulent cascade is determined by the hydrodynamics of the system:  MRI dynamo action is inhibited where sufficiently high resistivity is present.
\end{abstract}

\section{Introduction}
To fully understand planet formation, a global picture of protoplanetary disk (PPD) evolution is required, which implies understanding the interaction of magnetic field with partially ionized gases, or plasmas, often with significant amounts of dust. Magnetohydrodynamics allows the exploration of the planetary formation environment, turbulent angular momentum transport, interactions with the disk and, finally, orbital migration.
\newline
\newline
One of the main difficulties in understanding PPD dynamics lies in the mechanism(s) allowing accretion of material onto the star, which must remove angular momentum of the accreting material itself in orbital quasi-equilibrium, allowing flows into the inner regions of the disk, shaping the disk structure, as has been observed by ALMA. Possible sources of angular momentum transport are magneto-centrifugally driven winds (e.g. \citet{Blandford_Payne:1982}), and the effective viscous stresses introduced by the magnetorotational instability (MRI, \citet{Hawley_Balbus:1995}), whose nonlinear outcome in the ideal-MHD limit is the development of MHD turbulence. 
Self-gravity in conjunction with differential rotation has also been examined as a mechanism for driving turbulence by \citep{Wada_1999,Wada_2007} as well as the effect of hydrodynamical instabilities \citep{Zeldovich:1981}. Placing this model in the context of global disk structure and its interaction with the central star makes the problem very challenging, involving a wide range of temporal and spatial scales, coupled via nonlinear dynamical processes. 
%
%
%
%
\newline
\newline
Accurately capturing the full non-ideal physics is computationally difficult, and numerical expense limits the feasible resolution and/or run length, so global simulations but also local shearing box simulations have been carried out to study the saturation of the MRI ( \citet{Balbus_Hawley:2003}). Key aspects only addressed by global models are the actual transport of angular momentum, the wind launching, the feedback of magnetic fields on disk structure and the long-term evolution of the disk.
One of the important findings by local MHD simulations is that the net vertical magnetic field controls the saturation level of the turbulence (\citet{Hawley_Gammie_Balbus:1995,Sanoetal:2004,Pessahetal:2007,Suzuki_Initsuka:2009,Okuzumi_Hirose:2011,Simon_et_al:2013a,SImon:2018}),which essentially determines the strength of the transport of angular momentum and resulting mass accretion \citet{Suzuki_Initsuka:2014}.
In the shearing box the accretion is not actually simulated because of the symmetries which characterize the setup, while the shearing motion generates the effective viscosity (helping the angular momentum transport) through the MRI instability, so that the accretion rate is simply estimated from the stress tensor under the time-steady condition.
%
%
%
%
\newline
\newline
Except for the innermost regions of PPDs, where the temperature $T \sim1000K$, and the disk surface layers ionized by sources such as stellar X-rays, FUV photons and galactic cosmic rays, non-ideal MHD effects due to the low ionization levels of the gas (e.g. \citet{Blaes_Balbus:1994,Sano_et_al:2000}) are expected to be important. These processes are dominant across most radii in protoplanetary disks (\citet{Armitage:2011,Turner_et_al:2014}). \citet{Gammie:1996} proposed what has now become the traditional dead-zone model in which disk surface layers accrete by sustaining MRI turbulence, with the shielded interior maintaining an inert and magnetically decoupled dead zone. Here, MRI turbulence is quenched by competing non-ideal MHD terms, depending on density, temperature, degree of magnetization, the grain distribution and ionization (\citet{Balbus_Terquem:2001,Kunz_Balbus:2004,Desch_Turner:2015}), i.e. in the location within the disk.
In disk regions between $1-5$ AU, ohmic resistivity will be dominant near the mid-plane, the Hall effect at intermediate disk heights  (intermediate densities \citep{Wardle:2007}), and ambipolar diffusion (AD) in low density regions, higher up in the disk (e.g.\citet{Desch:2004}). 
\newline
\newline
Though such non-ideal effects have long been recognized (e.g. \citet{Sano_Stone:2002a,Sano_Stone:2002b}) and studied with an analytical approach \citet{Wardle:1999}, it is only recently that shearing box simulations including AD and the Hall term have begun to be performed in the relevant parameter regimes with significant resolutions, leading to a modified picture of how disks accrete that deviates significantly from the traditional dead zone (e.g. \citet{Sano_et_al:2000, Ilgner_Nelson:2006, Wardle:2012}).
\citet{Lesur:2014} included all three non-ideal MHD effects, and fond that if ${\Omega} \cdot {\bf B} > 0$, the Hall effect can produce an azimuthal magnetic field and so a large-scale Maxwell stress throughout the midplane of the disk. This result does not only make the disk more active in terms of accretion but can also increase the vertical scale hight of the disk. \citet{Bai:2014} showed such an amplification of the horizontal field at the midplane drives stronger winds and enhances the wind driven accretion up to 50 \%. \citet{Simon:2015}  remarked that the Hall effect is important even to qualitatively understand the disk structure and the accretion process and found bursty accretion events, possible due to Hall mediated whistler unstable modes in the disk. The role of non-ideal effects has also been studied in the context of global simulations \citep{Gressel_et_al:2015,Bethune:2017}, in comparison with shearing box models \citep{Bai_Stone:2013b}, showing the wind solution arises naturally in global simulations.
%
%
Before building a comprehensive model of global accretion disks it is necessary to understand the basic local properties of the plasma in which planet formation is embedded. 
\citet{Fleming_Stone:2003} studied, within a local shearing box setup, the evolution of MRI in vertically stratified accretion disks, i.e. with the ionization degree depending with height. They found the disk to remain quiescent in the central resistive region of the domain, while Reynolds stresses remain above $10\%$ the Maxwell stresses in the active layer, producing a significant contribution to the effective viscosity, $\alpha$. They suggested a residual mass inflow in the resistive layers from the active zone. \citet{Okuzumi_Hirose:2011} found out that the vertical structure is mainly affected by the vertical magnetic flux and the critical heights, the latter defining the atmosphere, the active and the dead zone, and it is insensitive to the details of the resistivity profile. 
\newline
\newline
Even simulations that are supposedly carried out within ideal MHD are actually affected by some form of numerical resistivity.
\citet{Sano_Initsuka_Miyama:1998} assuming an initial weak uniform magnetic field in the vertical direction, introduced definition of the effective \footnote{With '' effective'' we mean based on small scale turbulent fluctuations.} magnetic Reynolds number $\bar{R}_m=v^2_A/(\eta \Omega)$, where $v_A$ is the Alfv\'en speed, $\eta$ is the magnetic diffusivity and $\Omega$ the angular velocity. They obtain this magnetic Reynolds number $R_m= v_A L/\eta$ assuming $L=v_A/\Omega$, i.e. the characteristic MRI length scale. This allowed them to study the turbulence behavior in the nonlinear stage. They found that when $\bar{R}_m\ge 1$, the MRI does not saturate and channel flows develop in the system (see also \citet{Sano_Stone:2002a,Sano_Stone:2002b}).
In the case of a poloidal field with zero vertical net flux \citet{Fleming_Stone_Hawley:2000} argued that the MRI can be sustained when the effective magnetic Reynolds number, defined as $\bar{R}_m := c_s^2/\eta \Omega\ge10^4$. Notice the same relation can be written in terms of the Alfv\'en speed $v_A$, once the relation between the Alfv\'en and sound speed is established. They also defined a Reynolds number below which, in a numerical simulation with a typical vertical scale $H$ and other box size of length $L$, the computational box will be dominated by diffusion on time scale $\Omega$. They found this minimum Reynolds number to be $R^{cr}_m\sim (2\pi/L)(Hc_s/\Omega)$.
\citet{neumanBlackman:2017} defined a Reynolds number $R_m = L_x^2 \Omega/\eta $, where the macroscopic length scale $L_x$ is the size of the domain in the $x$ direction. They found a threshold value for the magnetic Reynolds number of $R^{th}_m \sim 1000$ for which the magnetic turbulence can be sustained. They also found that this findings are relevant to establish what are the numerical resistivity values which can guarantee the convergence of the MRI generated energy (kinetic and magnetic). We will discuss these concepts in the context of our own simulations in subsequent sections.
\newline
\newline
In this paper we investigate the basic properties of the local MHD turbulence set by the MRI in a shearing box setup, where the resistivity profile changes in the $x$ direction (radial direction in a global setup).
Differently from \citet{Okuzumi_Hirose:2011} we will consider a vertically uniform disk and we will address the effect of a vertical stratification in a future paper.  Notice that in the presence of a shear viscosity, with a vertical stratification, a meridional circulation pattern sets in in the poloidal plane of the disk because of the vertical gradient of the radial velocity \citep{Urpin:1984}. This results in a three dimensional transport within the disk height. 
In our case the disk is threaded with a vertical magnetic field with non zero net flux and we are interested in the radial transition region ($x$ direction in our simulations) between a resistive and an ideal zone. The goal of the paper is to understand the properties of MRI in such a region, which is considered of a paramount importance for the planetesimal formation. The paper is organized as follows: in Sec. \ref{sec:I} we discuss the shearing box concept and setup. In Sec. \ref{sec:II} we study the turbulence development and the effective viscosity in an ideal shearing box setup. We discuss the momentum equation balance and the spectral features of the MRI driven turbulence. In Sec. \ref{sec:C} we discuss the 1D spectral features in the shear direction, averaging in the vertical direction. In Sec. \ref{sec:III}  we investigate the turbulence development in a setup where the magnetic resistivity depends on the $x$ direction, with an ideal region and an area where the resistivity plays a role. In this context we study the origin of density accumulation and perturbations to the shear velocity at the transition between the resistive region and the ideal one. We discuss the spectral features of the transition region, comparing with the ideal simulation spectra. Finally we summarize our results in the conclusions. 

%
\section{ Setup for shearing box simulations.}
\label{sec:I}
%
%
The local shearing box approximation \citep{Stone_Gardiner:2010} adopts a frame of reference located at a radius $r_0$, corotating with the disk at orbital frequency $\Omega_{0} = \Omega(r_0)$. In this frame, the equations of resistive MHD are written in a Cartesian coordinate system (x, y, z)
\begin{eqnarray}
\label{eq:momentum_shearing}
&& \partial_t \rho + \nabla \cdot (\rho {\bf v})=0\\[1.5ex]
\label{eq:momentum_shearing1}
&& \partial_t (\rho {\bf v}) + \nabla \cdot (\rho {\bf v}{\bf v}+ {\bf \mathcal{T}})= \nonumber\\
&&\rho \Omega_{0}^2 (2qx\hat{i}- z\hat{k}) - 2 \Omega_{0}\hat{k} \times (\rho {\bf v})\\[1.5ex]
\label{eq:momentum_shearing2}
&& \partial_t {\bf B} = \nabla \times ({\bf v}\times {\bf B})- \nabla \times (\eta \nabla \times  {\bf B})
\label{eq:momentum_shearing3}
\end{eqnarray}
where $\hat{i}$, $\hat{j}$ and $\hat{k}$ are the unit vectors defining the orthonormal triad, and we assume the magnetic permeability to be unity.
We also adopt an isothermal equation of state $ P=\rho c_s^2$ and set $c_s=1$. 
The total stress tensor ${\bf \mathcal{T}}$ is defined as
\begin{equation}
{\bf \mathcal{T}}= (P+B^2/2) I- {\bf B}{\bf B},
\end{equation}
where $P$ is the gas pressure and $I$ is the unite tensor
\footnote{For the sake of clarity, we will use the capital letter $B$ or ${\bf B}$ to indicate the total magnetic field, and the lowcase $b$ or ${\bf b}$ to indicate the fluctuations. In the case there is no background field, e.g. in the $x$ direction $B_x\equiv b_x$.}.
An equilibrium solution for the set of equations \eqref{eq:momentum_shearing}-\eqref{eq:momentum_shearing2} is ${ \bf v}_0= -q \Omega_0 x \hat{j}$, where the shear parameter $q$ is defined as 
\begin{equation}
q= -\dfrac{1}{2}\dfrac{d \mathrm{ln} \Omega^2}{d \mathrm{ln}  r},
\end{equation}
i.e. for a Keplerian flow q= 3/2. The total velocity field is the equilibrium solution of the MHD equations, plus a perturbation,  ${\bf v} = {\bf v}_0 + \delta {\bf v}$.\\
The magnetic diffusivity $\eta$ depends in general on the location, and in particular in our model the profile $\eta(x)$ is described in Sec. \ref{sec:III}.
We also assume the disk is threaded by a constant, uniform vertical magnetic field ${\bf B}= B_{0z}\hat{k}=\sqrt{\dfrac{P_0}{2\beta_0}} \hat{k}$, where  $\beta_0$ is the plasma parameter at $t=0$, $P_0=\rho_0$ thanks to the isothermal equation of state, where $\rho_0$ is the initial uniform density, so  $\rho_0=1$. Since our box has no vertical stratification, magnetic field is not wound by the vertical differential rotation: being the temperature uniform, $\Omega=\Omega (x)$ does not depend on the vertical scale (generalization of the Von Zeipel theorem.\\
For our numerical calculations, we use the Athena++ code \citep{Stoneetal:2020}, a complete rewrite in C++ of the Athena code that integrates the shearing-box equations eq. \eqref{eq:momentum_shearing}-\eqref{eq:momentum_shearing3} using a standard Godunov scheme with second-order-accurate spatial reconstruction.
%
%
\section{\label{sec:II} Comparing ideal shearing box simulations.}
\label{sec:II}
%
%
In this section we discuss the turbulence development and force balance for ideal simulations described in Tab. \ref{tableSIM}, labelled with ID. 
We will then compare the latter with resistive setups (see Tab. \ref{tableSIM}, labelled with RES), to understand the effect of the size of the active zone.
For ideal simulations different runs have the same number of cells and physical parameters, but they differ in size. This corresponds to have different wave vectors ($k\sim2\pi n/L_z$, $n=(1,2,3...512)$ available for instabilities to grow. Still, the ratio between the Alfve\'n and sound speed with the maximum available shear decreases with larger boxes. 
Our fiducial model, labelled with IDB in Tab. \ref{tableSIM}, has $L_x=8$, $L_y=8$, $L_z=1$, since, as we will show later, this allows enough space to discuss the non uniform density accumulation in the resistive setups.\\
For all of the simulations we resolve the critical length scale of the MRI, $\lambda_C= 9.18 \beta^{-1/2} \sim 0.092$ \citep{Hawley_Gammie_Balbus:1995} and the maximum unstable wavelength of the MRI is $\lambda_{max}\simeq 2 \pi v_{A}/\Omega_0\sim 0.3$ (for $\beta=10^4$, see e.g. \citet{Suzuki_Initsuka:2010}), where all the lenghtscales are normalized to the scale height.
 \\[2ex] %
 \begin{table}
\begin{tabular} {|c|c|c|c|c|c|c|c|c|c|}
\hline
{\bf Name}              &$N_x$        & $N_y$       & $N_z$      &$\eta_0$ &$\beta$ &$L_x$ &$L_y $ &$L_z$ &$a$\\ 
\hline
IDA                         & $512$        &$512$          &$64$   &$0$&$10^4$      	&$ 4$ &$4$ &$1$ &$N/A$    \\
IDB                         &  $512$       & $512$         &$64$  &$0$&$10^4$     	&$ 8$ &$8$ &$1$ &$N/A$    \\
RESA                     & $512$        &$512$          &$64$     &$10^{-1}$  &$10^4$      	&$ 8$ &$8$ &$1$  &$0.1$  \\
RESB                   &  $512$       & $512$        &$64$      &$10^{-2}$  &$10^4$    		&$ 8$ &$8$ &$1$  &$0.1$   \\
RESC                   &  $512$       & $512$        &$64$      &$10^{-2}$  &$10^4$    		&$ 8$ &$8$ &$1$  &$0.01$   \\
\hline
\end{tabular}
\caption{Simulations parameters: N is the number of gridpoints in each direction, $\eta_0$ is the resistivity, $\beta$ is the plasma parameter, L is the simulation size in each direction (in unit of the vertical scale), a is the thickness of the transition between the resistive and the ideal zone, which does not apply (N/A) in ideal simulations.}
\label{tableSIM}
\end{table}
%
%
%
%
\subsection{\label{subsec:A} MRI development in ideal MHD shearing box simulations.}
\label{subsec:A}
\begin{figure}
 \includegraphics[width=90mm]{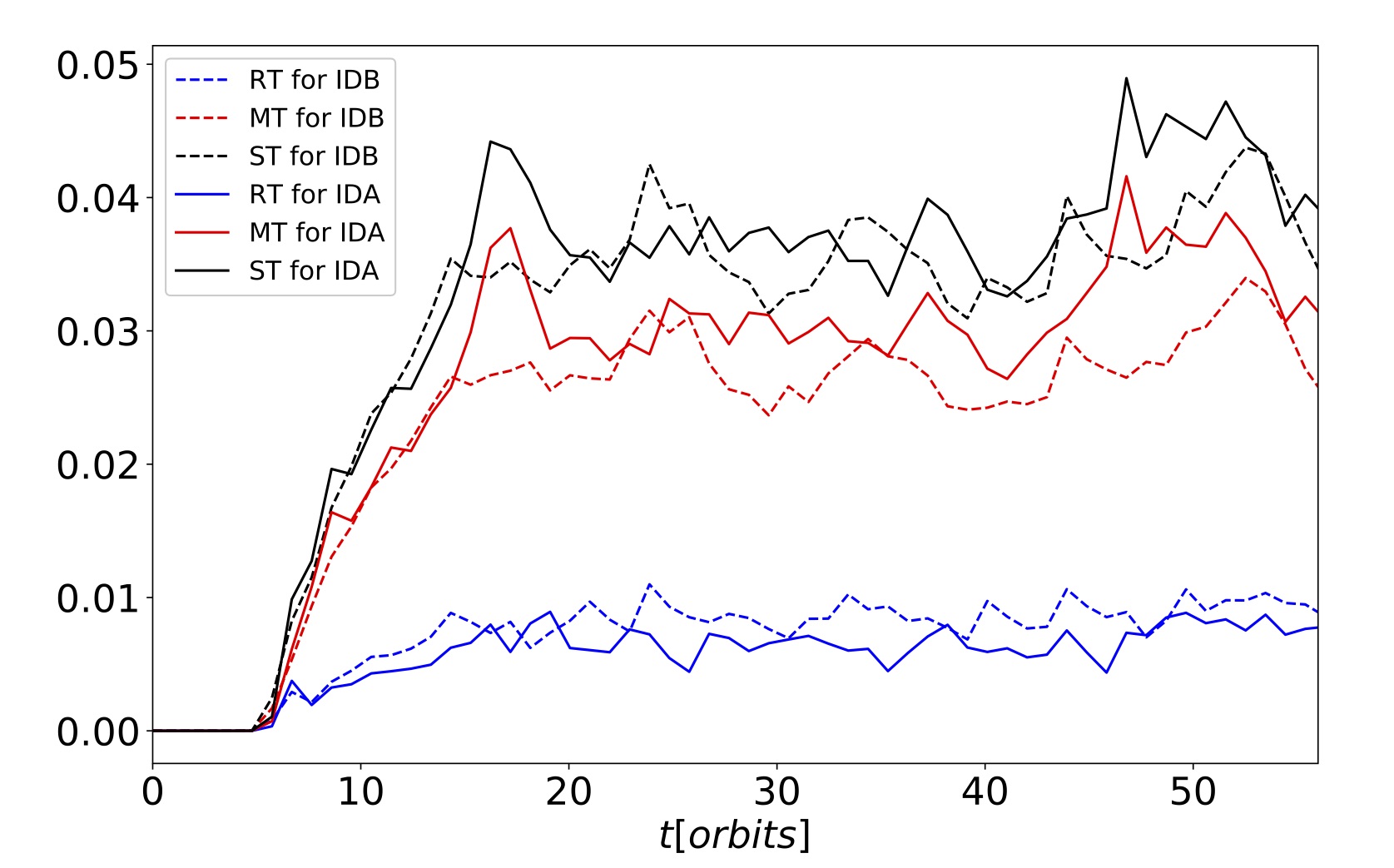}
\caption{Reynolds (RT, blue), Maxwell (MT, red) and Stress (ST, black) tensors for IDA (solid lines) and IDB (dashed lines), normalized to the pressure $P$. The saturation level for the stress tensor is around 0.035 for both simulations.}
\label{tensor_ideal}
\end{figure}

%
%
We quantify the efficiency of the turbulence through the $x-y$ component of the total stress tensor
\begin{eqnarray}
\label{stress_tensor}
T_{xy}=M_{xy}+R_{xy},
\end{eqnarray}
where, being $B_i$ and $v_i$ where $i=x,y$ the component of the magnetic and velocity fields respectively,  $M_{xy}=\langle-B_xB_y\rangle$ is the Maxwell tensor and $R_{xy}=\langle\rho v_x \delta v_y\rangle$ is the Reynolds tensor; the brackets indicate the average over the $y$, $z$ (vertical) direction.
In Fig.\ref{tensor_ideal} we show for simulation IDA (solid lines) the effective viscosity $\alpha=\langle T_{xy}\rangle/\langle P(t)\rangle$, where the average is over the whole volume, i.e. the stress tensor normalized with the average pressure $\langle P(t)\rangle$; we also show the breakdown in the Reynolds and Maxwell tensors, also normalized with $\langle P(t)\rangle$.
The saturation level of the stress tensor is about 0.035 in a case of a box characterized by a resolution of 64 grid points in the vertical direction , which is compatible with previous literature (e.g. \citet{Hawley:1996, Hawley:1995} and for a more recent simulation see e.g. \citet{Shi_et_al:2016}). 
As expected, the main contribution to effective viscosity $\alpha$ is due to the Maxwell tensor.
\\[2ex]
For simulation IDB, we expect to have convergence of the stress tensor to the same value, as the resolution in the vertical direction (shown to be a key feature for convergence by \citet{Hawley:1995}), is the same. Indeed this is what we observe in Fig. \ref{tensor_ideal} (dashed lines), where the relative contribution of the Maxwell and Reynolds tensor for IDA and IDB are similar.
\subsection{Numerical resistivity for ''ideal simulations''.}
Even if there is no explicit resistivity, we can estimate a magnetic diffusivity $\eta_N= v_A\,\Delta x=0.0002$, where in our simulations $\Delta x= 1/64=0.0156$. The macroscopic Reynolds number $R=v_A L/\eta=64$, where $v_A$ is the Alfv\'en speed based on the initial vertical magnetic field, $L=1$ is the size of the box (in unit of the vertical scale). The (numerical) magnetic Prandtl number, since the numerical diffusivity and viscosity are calculated in the same way, is of order 1. We expect the MRI to develop in regions where the macroscopic Reynolds number is $R_m>1$. 
In comparison with other definition of the critical Reynolds number (see the introduction), we obtained $R_m = L_x^2 \Omega/\eta = 32 \times 10^{4}$ for the parameter defined in \citet{neumanBlackman:2017}, which confirms magnetic turbulence should be sustained. Considering a similar definition by \citet{Fleming_Stone_Hawley:2000},  our $\bar{R}_m \sim 10^4$, which is the threshold value for the turbulence to be sustained. In terms of the nonlinear evolution of the turbulence, using the parameter defined by \citet{Sano_Initsuka_Miyama:1998}, we get $\bar{R}_m=v_A^2/\eta\Omega<1$, i.e. we do not expect channel flows to dominate the simulation.
%
%
\subsection{\label{subsec:B} Force balance for the saturated stationary state.}
\begin{figure}
\includegraphics[width=87mm]{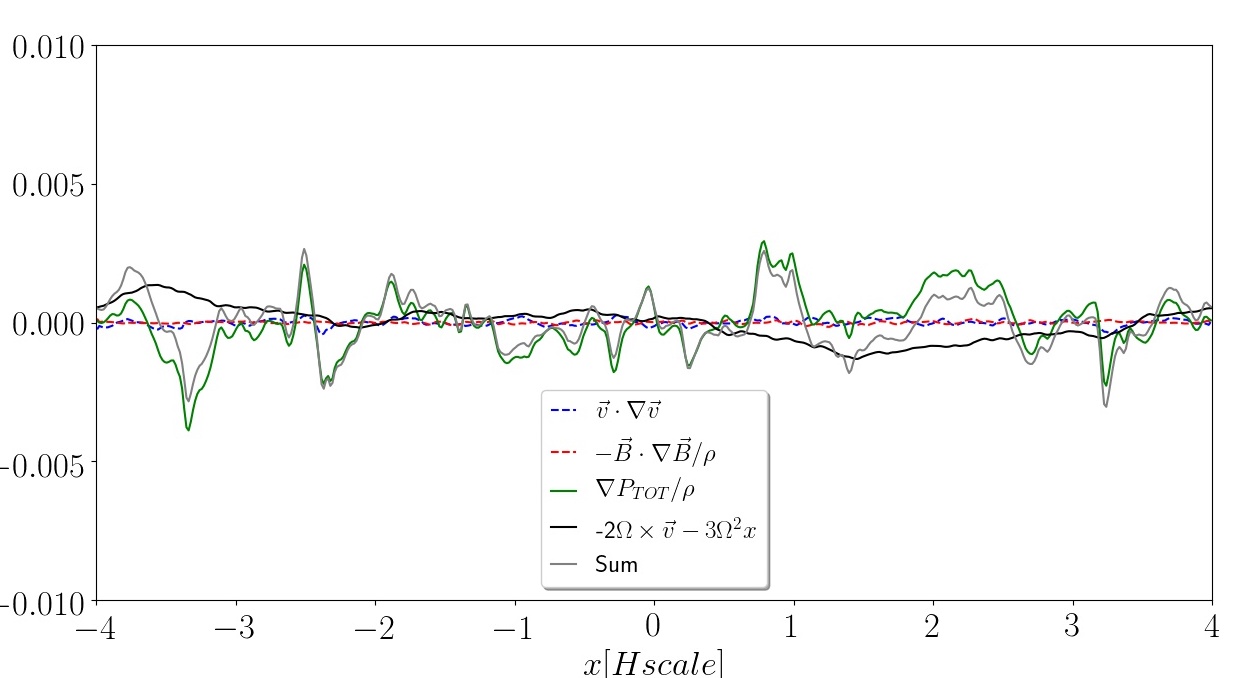}
\caption{Contribution of each term in the x-direction (radial) of Eq. \eqref{eq:momentum_shearing1}, i.e. the force balance for IDB. The fictitious forces balance each other and the residual difference between the Coriolis and the centrifugal force balances the fluctuations of the other terms. Each term is normalized the the maximum of the centrifugal force.}
\label{forcebalance-partial}
\end{figure}

%
%
As discussed in Sec.\ref{sec:I} a shearing box in a corotating frame with the disk includes the Coriolis and centrifugal terms in the momentum equation, see Eq. \eqref{eq:momentum_shearing1}. In Fig.  \ref{forcebalance-partial} we show the contribution of each term in Eq. \eqref{eq:momentum_shearing1} for our fiducial model IDB. We can see the total pressure gradient fluctuations are balanced by a perturbation of the total fictitious forces. The pressure gradient modification (or equivalently the density gradient, given our isothermal ansatz) generated by the changes in the Coriolis force, reflects the compressibility of the system. 
%
The fluctuations in the pressure gradient are evident at all scales, mainly reflecting the spatial fluctuations of the hydrodynamic pressure.

\section{Spectral features of MHD turbulence.}
\label{sec:C}

One of the goals of this study is to understand the physics at the transition between the ideal MHD and strongly resistive MHD domains. We begin by discussing one dimensional spectra in the azimuthal plane for the ideal shearing box simulations.
As we are looking for structures in the x-y plane (assuming the vertical direction is uniform), we will calculate the one dimensional Fourier transform along the y direction, for a selected position $x_1$. We will then calculate the power spectrum averaged over the vertical direction ($z$-direction). Finally we averaged over about 20 orbital times, once the MRI is saturated, to obtain the plotted quantity. In formulae, for any field component $A_i (x,y,z,t)$, defining averages $<>_x$ in terms of the subscript independent variable $x$, we have

\begin{eqnarray}
\label{eq:spectra}
&&\langle \rvert A_i(x_1,k_y)\rvert ^2 \rangle_{z,t}= \\[2ex]
\label{eq:spectra2}
&& \langle \biggr\rvert \int {A}_i(x_1,y,t) e^{-ik_y\, y} dy \biggr\rvert ^2 \rangle_{z,t}\\[2ex]
\label{eq:spectra3}
&& k_y = 2\pi n/L_y
\end{eqnarray}
with $n=(1,2,3...512)$. \\[1.5ex]
\subsection{\label{subsec:C}Ideal MHD spectra.}
\begin{figure}
 \includegraphics[width=90mm]{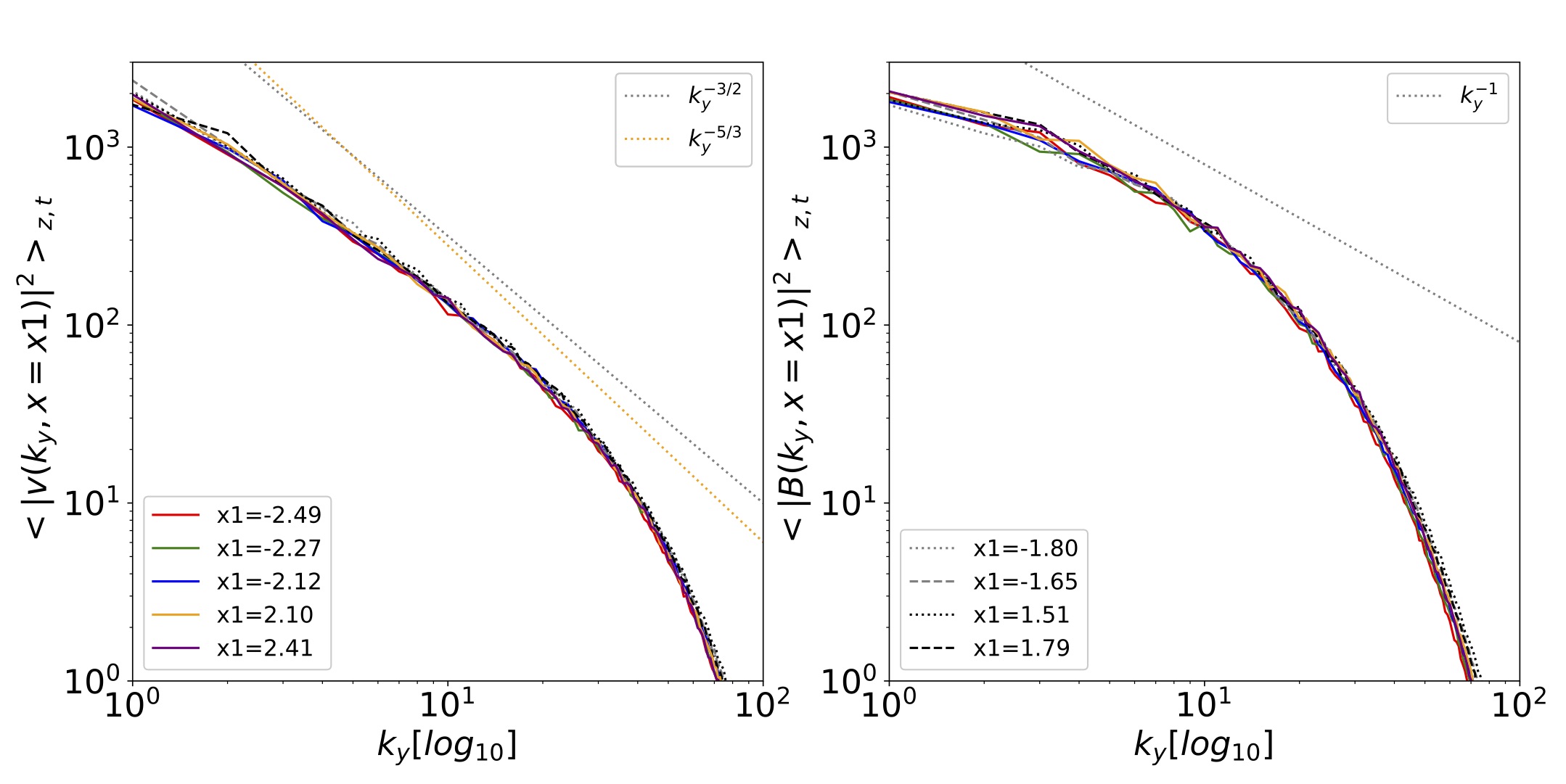}
\caption{Kinetic (left) and magnetic (right) energy 1D spectra, defined in Eq. \eqref{eq:spectra}, for simulation IDB. The spectra are averaged in the vertical direction (z) and over 20 orbital times, between orbit $40$ and $60$. Color labels the location x1 at which the one dimensional Fourier Transform has been calculated. For a better visualization the legend is spread among the two panels and refers to both of them. The velocity spectral slope is closer to -3/2, compared to the yellow dashed line indicating the Kolmogorov slope -5/3.}
 \label{spectra1}
\end{figure}

The result for the velocity field (RMS) is plotted in Fig. \ref{spectra1} (left), for the ideal simulation IDB, where colors label different values $x_1$.
Fig. \ref{spectra1} (left) shows that a powerlaw can be identified in the kinetic energy spectrum. Fitting the points between $k_y=2$ and $k_y=20$ the velocity field spectral slope resulted to be close to $-3/2$. We also plotted the -5/3 slope as reference.
Fig. \ref{spectra1} (right) shows the magnetic energy spectrum at MRI saturation, where a powerlaw is much harder to identify.
In addition, the MRI generated turbulence is not strongly magnetized, and the plasma $\beta$ is very large. Intriguingly, solar wind turbulence, at a plasma $\beta\sim1$, also shows velocity field spectral slopes close to $-3/2$, flatter than magnetic field spectra (see \citet{BrunoCarbone:2013}) in the inertial range, but steeper than magnetic energy spectra at the largest scales, where the powerlaw in the solar wind is closer to $k^{-1}$. Our magnetic spectra have energies comparable to the velocity field at large scale, where the magnetic spectrum also appears to be relatively flat. However, the powerlaw is visible for less than one decade, and it seems clear that injection is dominating at large scales. The spectra fall off at values close to $n\sim70$, consistent with the magnetic Reynolds numbers estimate given above. We would like to remark that larger magnetic Reynolds numbers should allow more extended inertial range and accordingly to the results in \citet{neumanBlackman:2017}, to better resolve the turbulence, finding higher saturation values.
%
%
%
%
\section{\label{sec:III} MRI development in shearing box simulations with a non uniform resistivity profile.}
\label{sec:III}
\begin{figure}
 \includegraphics[width=90mm]{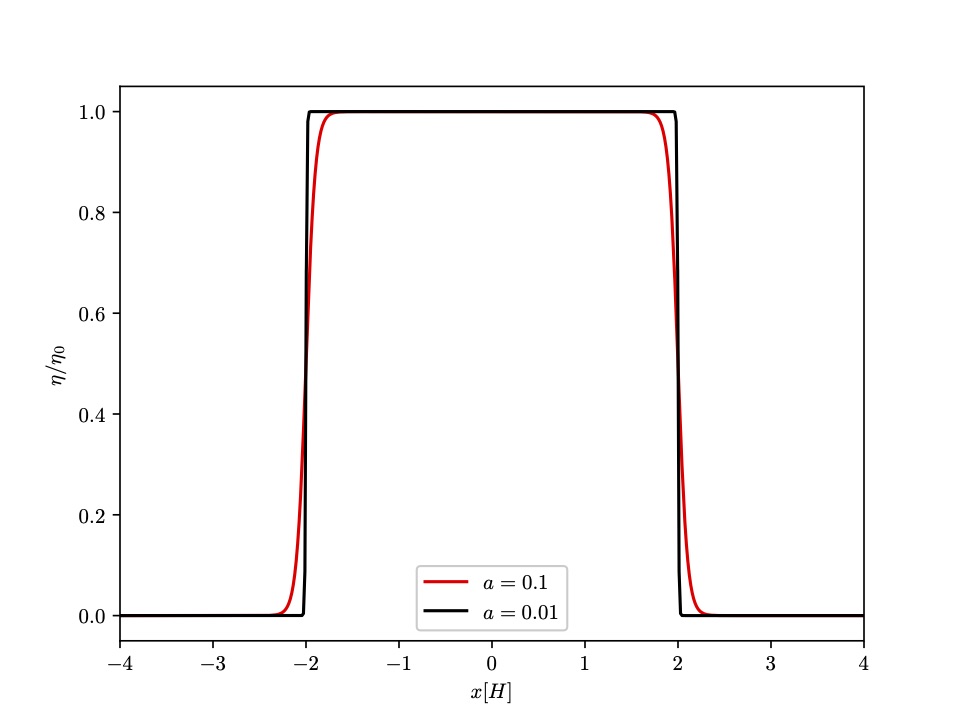}
\caption{Resistivity profile as a function of the x coordinate, for $x_0=1$ and $a=0.1$ (red) and $a=0.01$ (black).}
 \label{res_profile}
\end{figure}

%
%
For resistive simulations the ohmic resistivity profile is shown in Fig.\ref{res_profile} and described by:
\begin{equation}
\label{resistivity}
\eta(x)=\dfrac{\eta_0}{2}(\mathrm{tanh}((x+x_0)/a)-\mathrm{tanh}((x-x_0)/a)),
\end{equation}
where $x_0=2$.
The set of simulations we performed are described in Tab. \ref{tableSIM} and labelled with RES.
%
%
%
RESA and RESB differ for the value of the resistivity $\eta_0$, while in RESC the parameters are the same as RESB but the transition region is 10 times thinner. In RESB the transition is resolved by 6 grid points while in RESA the transition is not resolved.
Using for example the explicit resistivity for RESB, we can estimate the macroscopic Reynolds number $R=v_A L/\eta=0.71<1$, so we expect the MRI to be inhibited in the regions where the macroscopic Reynolds number is $R_m<1$. 
The MRI modes should be then quenched so it is worth it to compare with the definition in \citet{Fleming_Stone_Hawley:2000}, of  a Reynolds number $R^{cr}_m\sim (2\pi/L)(Hc_s/\Omega)$ below which, the computational box will be dominated by diffusion on time scale $\Omega$. In our simulation $R^{cr}_m=0.63$ so in the dead zone $R_m \sim R^{cr}_m$, i.e. based on this criterion all the MRI modes are damped. Please notice this is even more relevant for RESA for which the explicit resistivity is higher.

\begin{figure}
\centering    
\subfigure{\label{fig:a}\includegraphics[width=90mm]{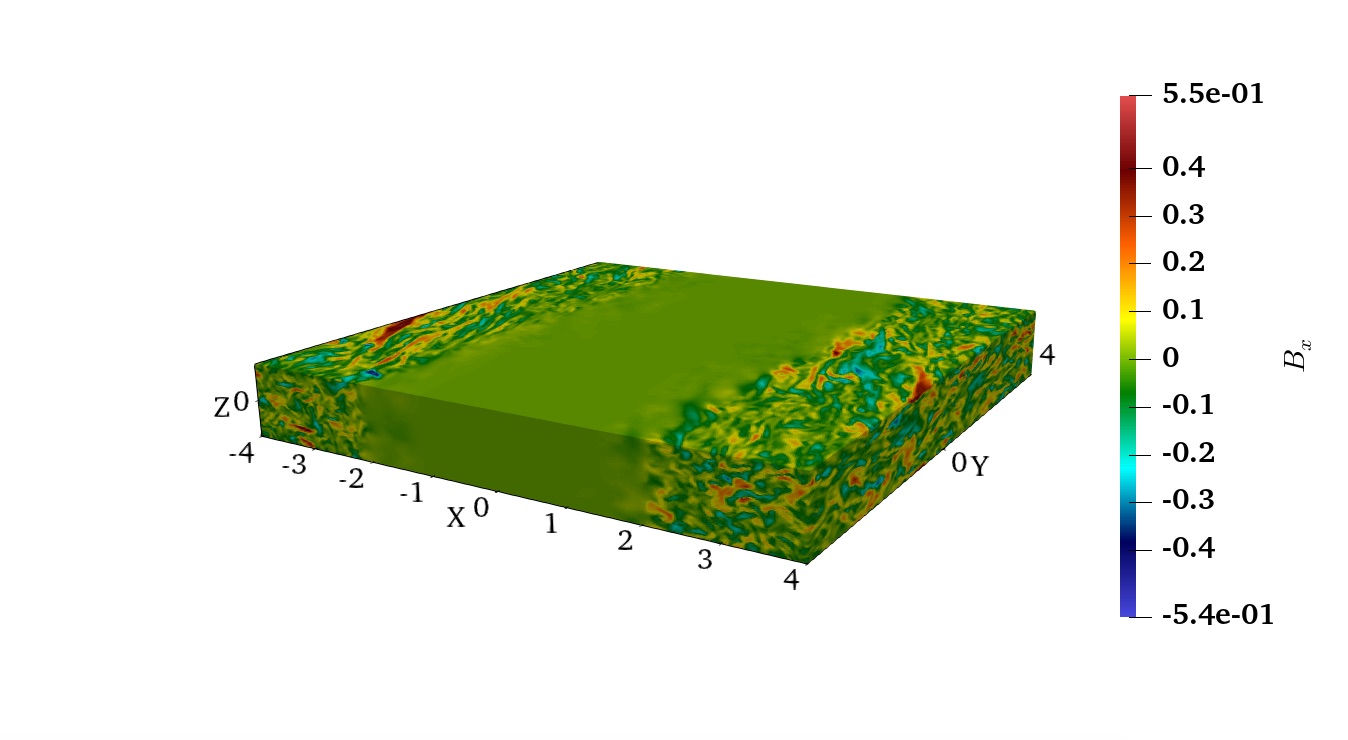}}
\par\smallskip
\subfigure{\label{fig:b}\includegraphics[width=90mm]{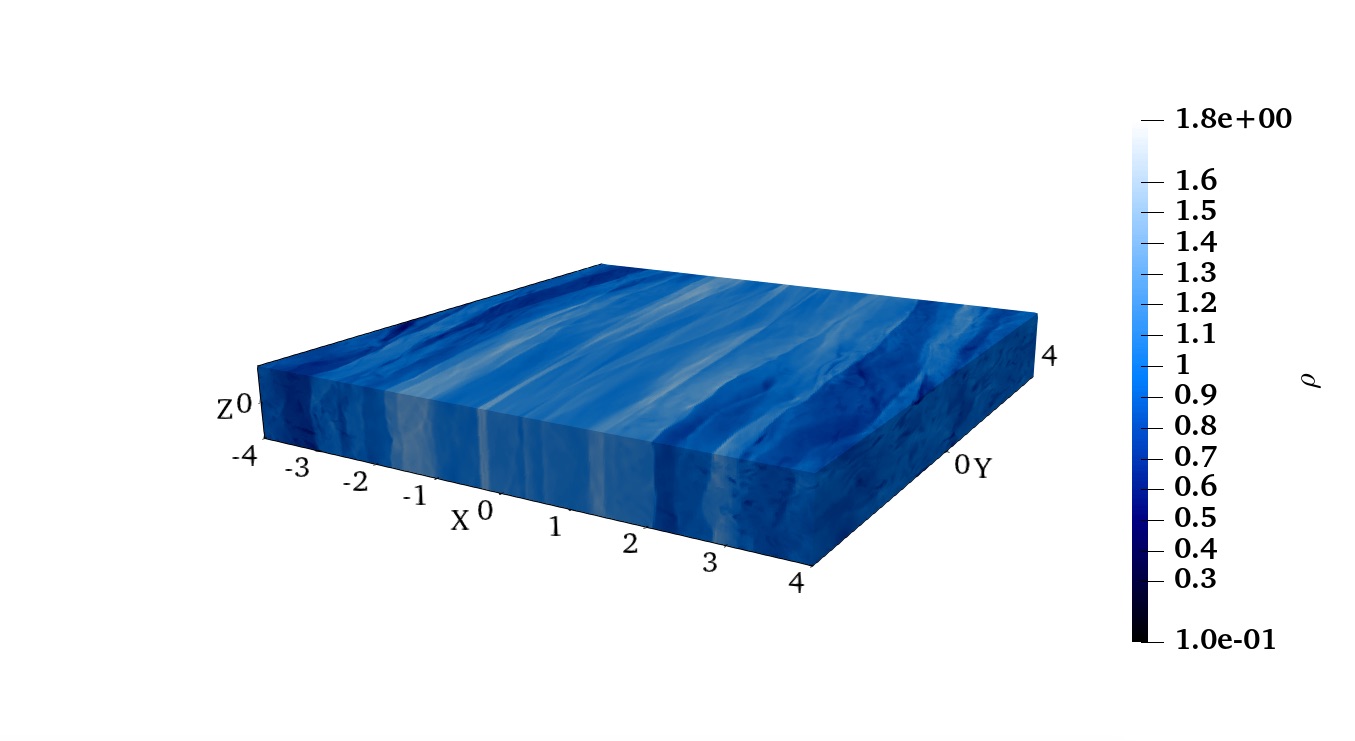}}
\caption{3D profiles in simulation RESB after $\sim 20 $ orbital times for (top) x component of the magnetic field, where we see the field is developed in the ideal region of the domain and the transition to the inactive area is not sharp. (bottom) Density profile $\rho$. Density fluctuations are oblique in the $x-y$ plane.}
\label{Figure:3D}
\end{figure}

In Fig. \ref{Figure:3D} (a) we show an example of how the density and magnetic field look like in our simulation RESC after $\sim 20 $ orbital times, when density seems to accumulate in the central resistive region (see the discussion in Sect.\ref{subsect:BI}).  A turbulent magnetic field develops in the $x$ and $y$ direction in the active zone, while in the resistive zone the magnetic turbulence is quenched. Oblique density fluctuations in the $y-z$ plane are present, see Fig. \ref{Figure:3D} (b).
%
%


\subsection{\label{subsec:AI} Stress tensor in a non uniform resistivity setup.}
\begin{figure}
\hspace{-0.9cm}
 \includegraphics[width=100mm,left]{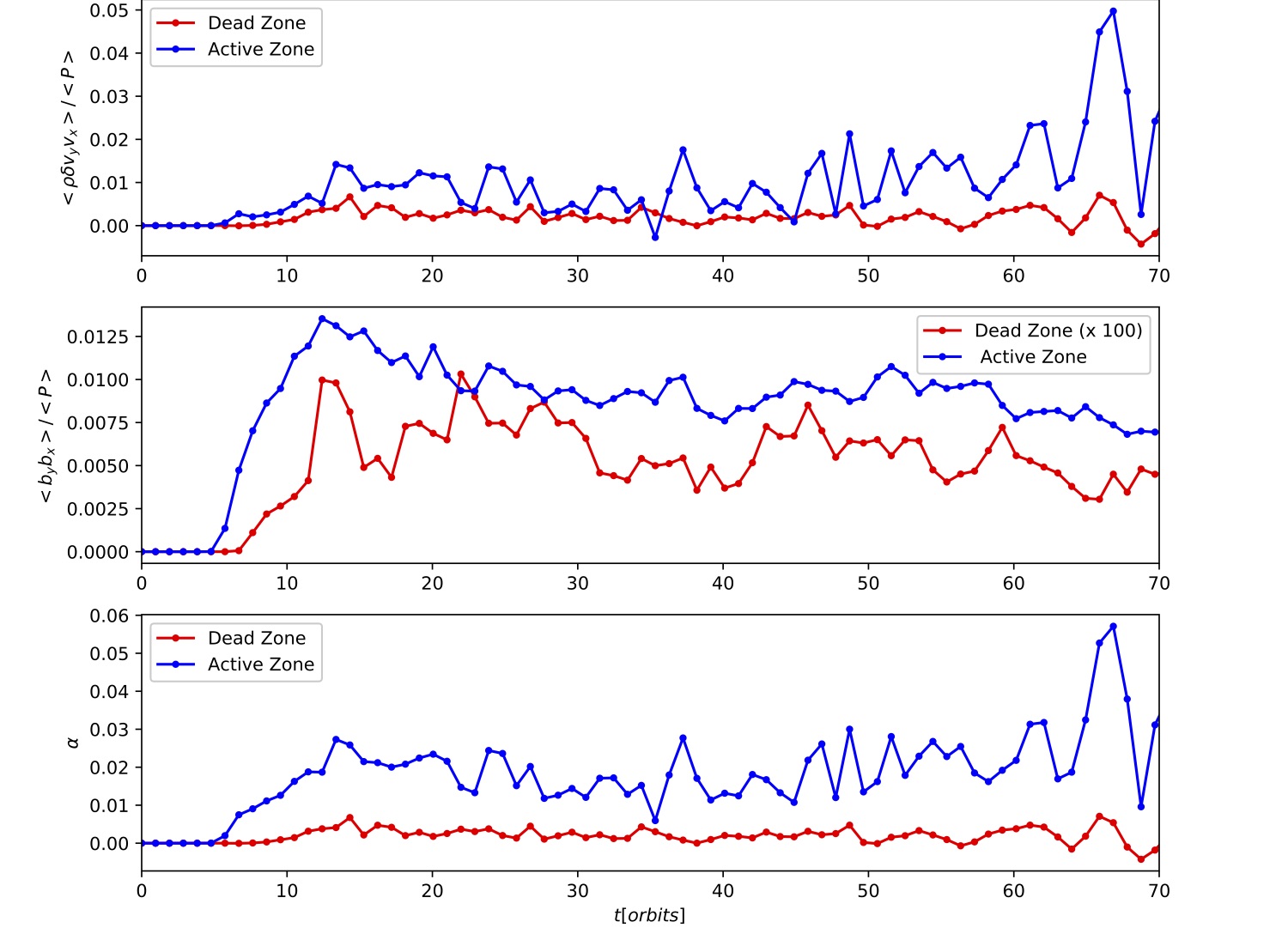}
\caption{(top) Reynolds tensor, (center) Maxwell tensor, (bottom) Stress tensor for simulation RESB, normalized with the average pressure $P$ at each time step.}
\label{tensor_resistiveB}
\end{figure}

Our goal is to compare the structure and distribution of the turbulence in the active and dead zones with particular interest in the boundary between the dead and active regions.
Since, as shown in Fig.\ref{tensor_ideal}, the main contribution to the effective viscosity $\alpha$ is due to the magnetic tensor, we expect to see a significant variation in the case of a resistive domain. 
In Fig.\ref{tensor_resistiveB}, for our fiducial simulation RESB, we show the (top) Reynolds tensor, (center) Maxwell tensor, normalized to the pressure $P$, and how they contribute to the total (bottom) $\alpha$ (Stress tensor normalized with the pressure $P$) averaged in z and y direction, as defined in Eq. \eqref{stress_tensor} and relative description in Sec.\ref{subsec:A}.
Each panel shows an average of the ideal regions (green) and the resistive central region (red).

For the first 50 orbits the stresses in the ideal region are quite similar (in terms of saturation levels and trends) to our fiducial model, simulation IDB, see Fig.\ref{tensor_resistiveB}, even if the actual saturation level in RESB is slightly less than 0.03. This lower value with respect to the ideal case is due to the lack of activity in the resistive region, which is suppressed by a relatively high resistivity. Indeed, we report for RESA, for which the resistivity is even higher $\eta_0=0.1$, a saturation level of 0.01. \\
Between $t\sim 20$ and $t\sim50$ it seems the MRI reached a saturated state. 
After $t\sim 50\ orbits$, very high fluctuations in the stress tensor appear, making its contribution dominant for the effective viscosity.  After $t\sim 60 orbits$ the Maxwell tensor grows again, most probably due to slow growing MRI modes. Indeed density and magnetic waves (with smaller amplitude) are excited at the transition region, due to the energy injection from the active region. \\[2ex]

\subsection{Comparing different resistivity setups.}
\begin{figure}
\hspace{-0.4cm}
\subfigure{\includegraphics[width=95mm]{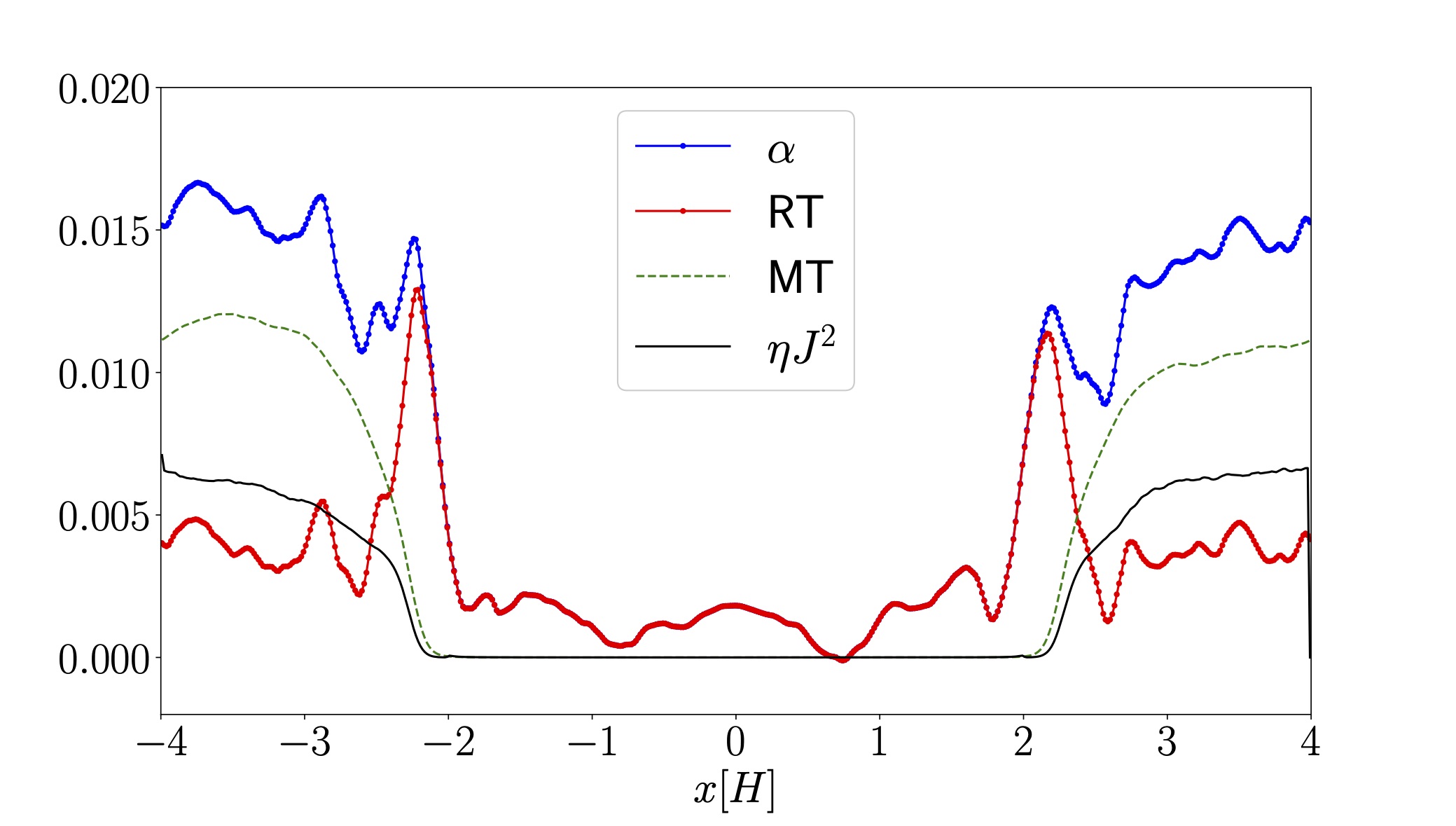}}
\par\smallskip
\vspace{-0.4cm}
\hspace{-0.4cm}
\subfigure{\includegraphics[width=95mm]{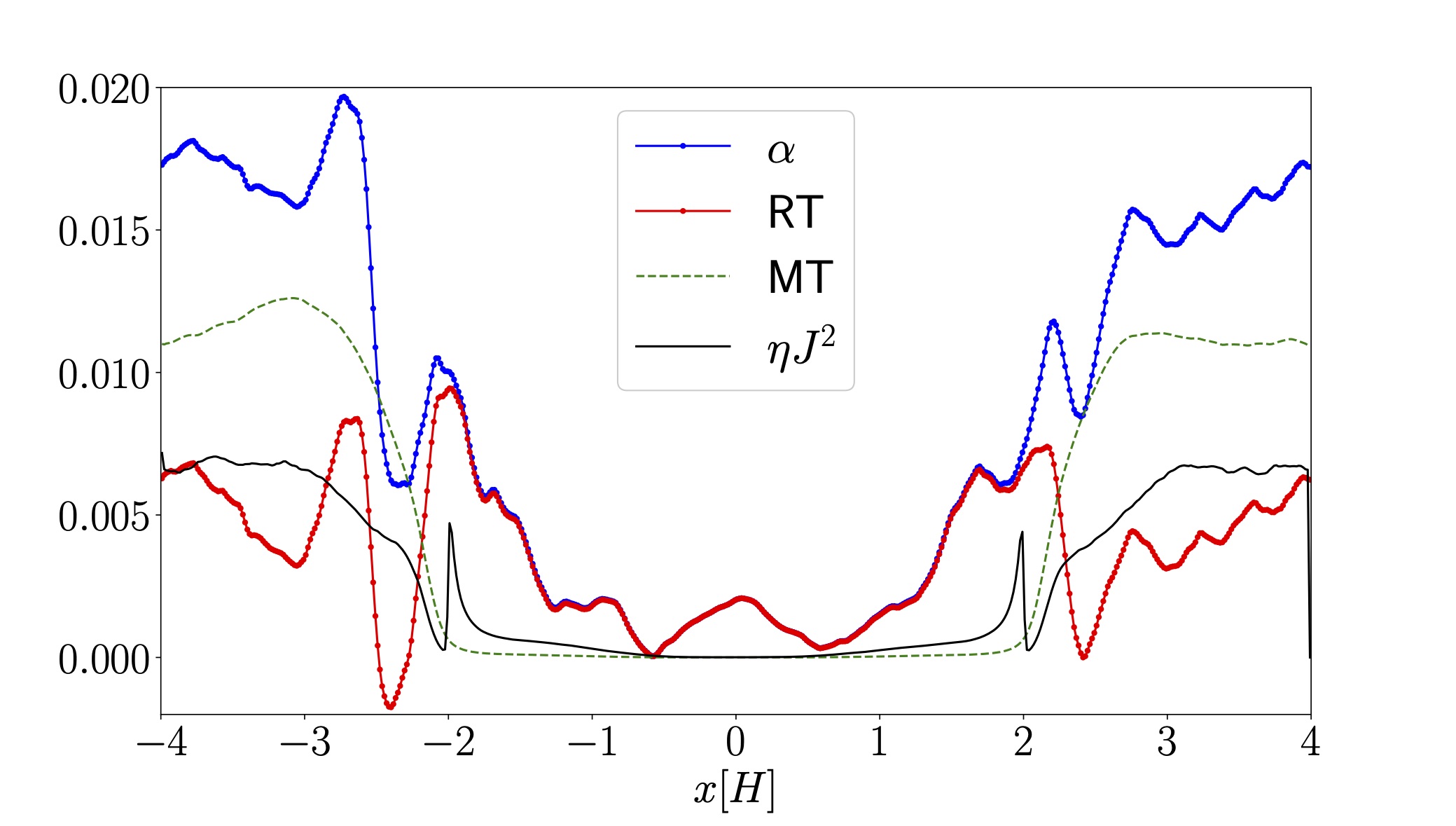}}
\par\smallskip
\vspace{-0.4cm}
\hspace{-0.4cm}
\subfigure{\includegraphics[width=95mm]{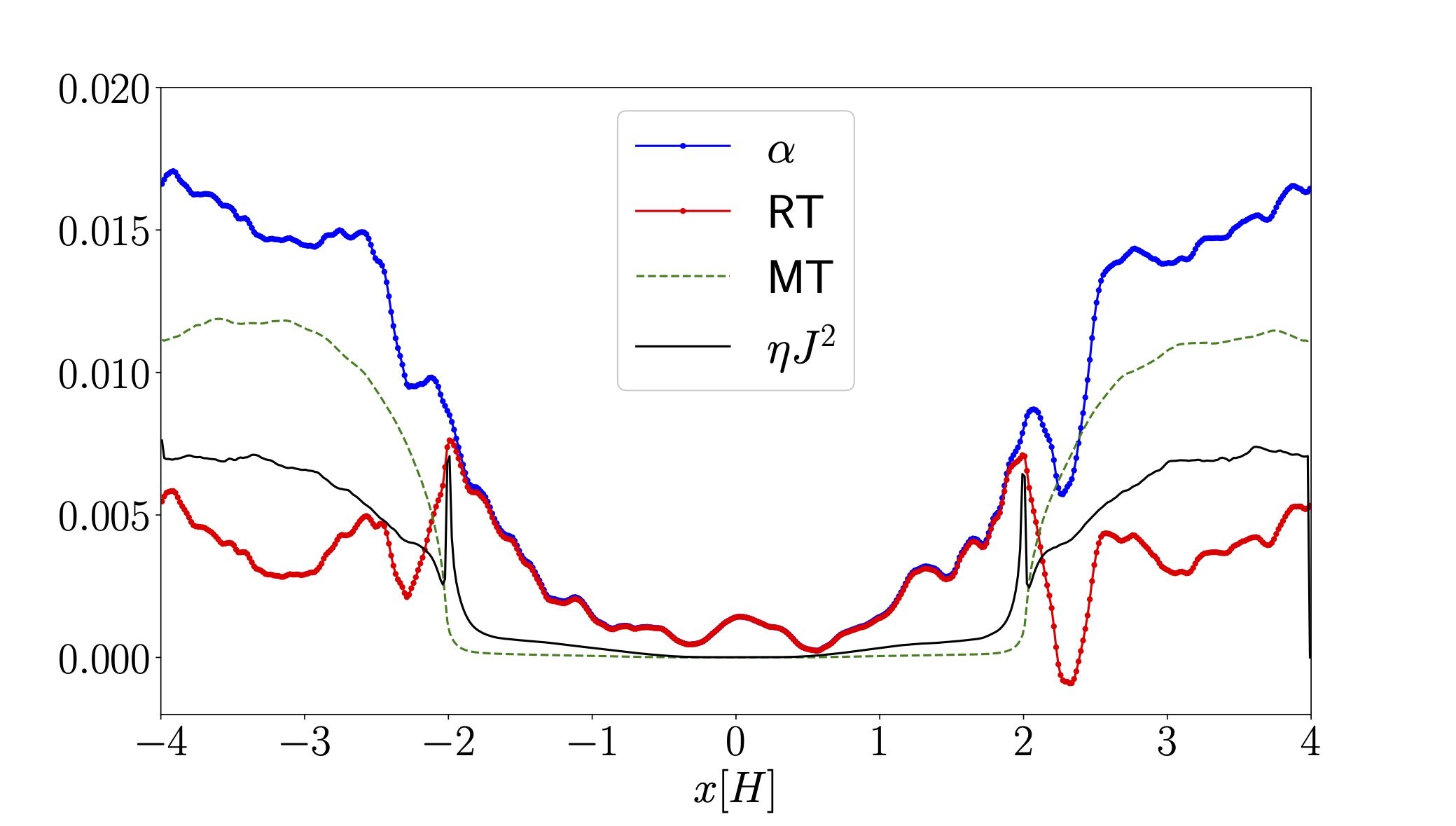}}
\caption{Effective viscosity $\alpha= \langle T_{xy}(t)/P(t)\rangle_{y,z,t}$ and breakdown in the hydrodynamical stresses $ \langle R_{xy}(t)/P(t)\rangle_{y,z,t}$ and magnetic stresses $ \langle M_{xy}(t)/P(t)\rangle_{y,z,t}$ for simulation RESA (top), RESB (center) and RESC(bottom), where $P(t)$ is the pressure at each point and time step and $T_{xy}$ is defined in Eq.\ref{stress_tensor}.
The spatial average is in the y-z plane and the average in time is over 20 orbits in the saturated phase (25-45 orbital times). The black solid line shows the magnetic dissipation $\eta |{\bf J}|^2$, see text for further explanation.}
\label{alpha}
\end{figure}

In order to understand the dependence of each quantities on the distance from the transition region, in Fig. \ref{alpha} we show the effective viscosity $\alpha$ as a function of $x$ averaged in the $y=z$ plane for RESA (top), RESB (center) and RESC (bottom). The quantities are averaged over 20 orbital times, in the saturated state. It is clear that the turbulence is sustained through the transition region and in the resistive region in RESB and RESC even if the average value of $\alpha$ in the resistive region is less than $20\%$ of the value in the ideal zone for both cases. The ten times higher resistivity value in simulation RESA, produces a sharper transition in the magnetic turbulent activity, reflecting in lower values of $\alpha$. In all of the three setups the Reynolds tensor is highly fluctuating, keeping the effective viscosity $\alpha$ relatively high even in simulation RESA. 
The sharper transition for RESC is slightly visible at $|x|=2$ where the magnetic stress tensor falls from $5\times 10^{-3}$ to 0. The Reynolds, and so the resulting total Stress tensor, do not differ significantly in sim RESB and RESC because the feedback on the velocity due to the sharper resistivity transition, occur through the magnetic fluctuations, appearing as quite smooth in both cases. From this analysis, it emerges the most important parameter to determine the turbulence behavior is the actual value of the resistivity in the dead zone, while the thickness of the transition region does not significantly affect the turbulence values.

\subsection{\label{subsec:AII} Magnetic dissipation.}

We checked the magnetic dissipation integrated over the volume to better understand the resistive case stress tensor saturation level. We calculated $|{\bf J}|^2=|\nabla \times {\bf B}|^2$ then we multiplied by the numerical resistivity $\eta_N\sim 0.0002$; in the resistive cases we adopted the resistivity defined in eq. \eqref{resistivity} plus the numerical resistivity value $\eta_N$. The results are plotted in Fig. \ref{alpha} (solid black line). We found the magnetic dissipation to be significantly stronger in the ideal region, and in general in ideal simulations IDB. This suggests the magnetic flux penetration in the dead zone occur only in the layers closer to the active area and limits the possibility to dissipate magnetic field further inside the resistive region, eventually leading to plasma heating when the feedback on temperature is taken into account, reflecting in changes in the ionization degree only in the transition region. In Fig. \ref{alpha} (top) we can see in simulation RESA for $|x|<2$ the dissipation is indeed zero. In simulation RESB instead, the dissipation is zero only in $|x|<0.5$, suggesting a deeper penetration of the turbulent magnetic field in the resistive region. In Fig. \ref{alpha} (bottom, solid black line, we show the dissipation for run RESC, where the transition in the dissipation is very sharp. Even if not resolved, it allows us to conclude that the turbulence penetration through the boundary of the dead zone region is determined by the value of the resistivity itself, through the magnetic dissipation process.


\subsection{\label{subsect:BI} Density accumulation and streams at the transition from the dead to the active zone.}
\label{subsect:BI}
 
\begin{figure}
 \includegraphics[width=90mm]{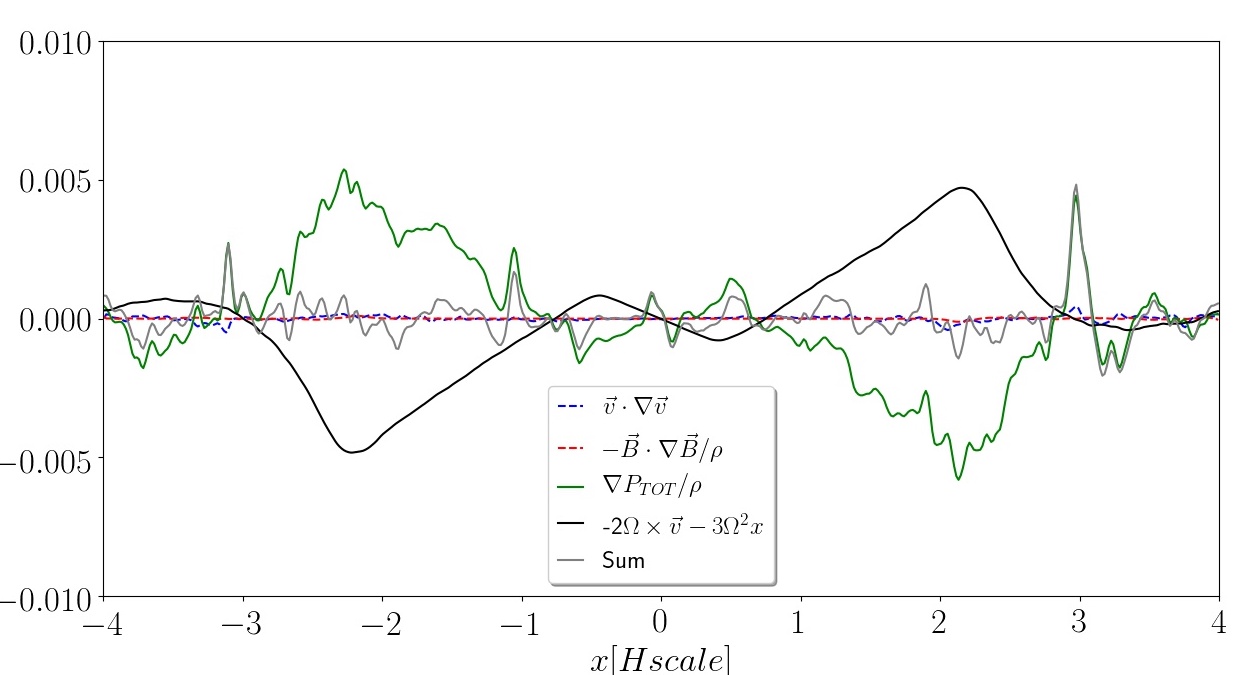}
\caption{Momentum equation Eq. \eqref{eq:momentum_shearing} balance for RESB. The total pressure tensor (green solid line) balances the fictitious forces, i.e. Coriolis and Centripetal force (black solid line). The other terms present small fluctuations. Each term is normalized to the maximum of the centrifugal force.}
 \label{forcebalance3}
\end{figure}

\begin{figure}
\hspace{-0.6cm}
 \includegraphics[width=103mm]{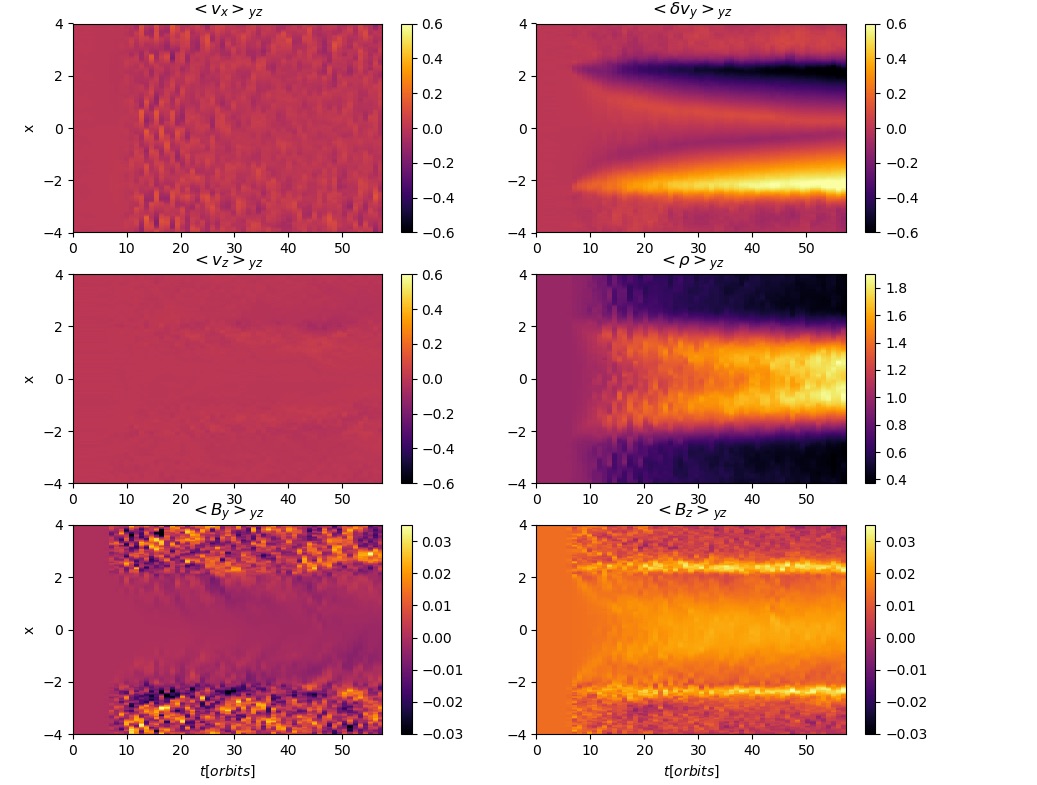}
\caption{2D profiles of the fields, averaged in the $y,z$ directions for different $x$ (vertical axis), as a function of time (horizontal axis), for simulation RESB. On the left column: (top) x and (center) z components of the velocity field, (bottom) y component of the magnetic field (generated by the MRI). The right column shows: (top) y component of the turbulent velocity $\delta v_y(t)=(v_y(t)-v_0)_{max}$, (center) density, (bottom) z component of the magnetic field. 
It is evident there is a temporal and spatial correspondence of the features characterizing the quantities in the right column, 
suggesting a common physical process originating them.}
 \label{2DProfilesRESB}
\end{figure}

One of the characteristic features of the resistive setup is the presence of density peaks in the dead zone or, as it appear after a more detailed analysis, in the proximity of the transition regions. In Fig.\ref{2DProfilesRESB} we show the variation of the profiles in the $x$ direction as a function of time for simulation RESB. 
The density peak starts forming after $\sim 6$ orbits with MRI kicking in (see corresponding stress tensors), in correspondence with the formation of an additional velocity component in the $y$ direction, at the transitions between the ideal and the dead zone. The density peak reaches $\rho=1.6 \rho_0$. 
In Fig. \ref{forcebalance3} we show each contribution to the $x$ component of Eq. \eqref{eq:momentum_shearing1}, where it is clear that the balance for the fluid pressure term comes from the Coriolis force. Since we employ an isothermal equation, the density has the same role as the pressure in the force balance, i.e. its profile is altered by the changing in the Coriolis force in the nonuniform resistivity setup of run RESB.
\subsection{MRI density structures.}
\begin{figure}
 \includegraphics[width=90mm]{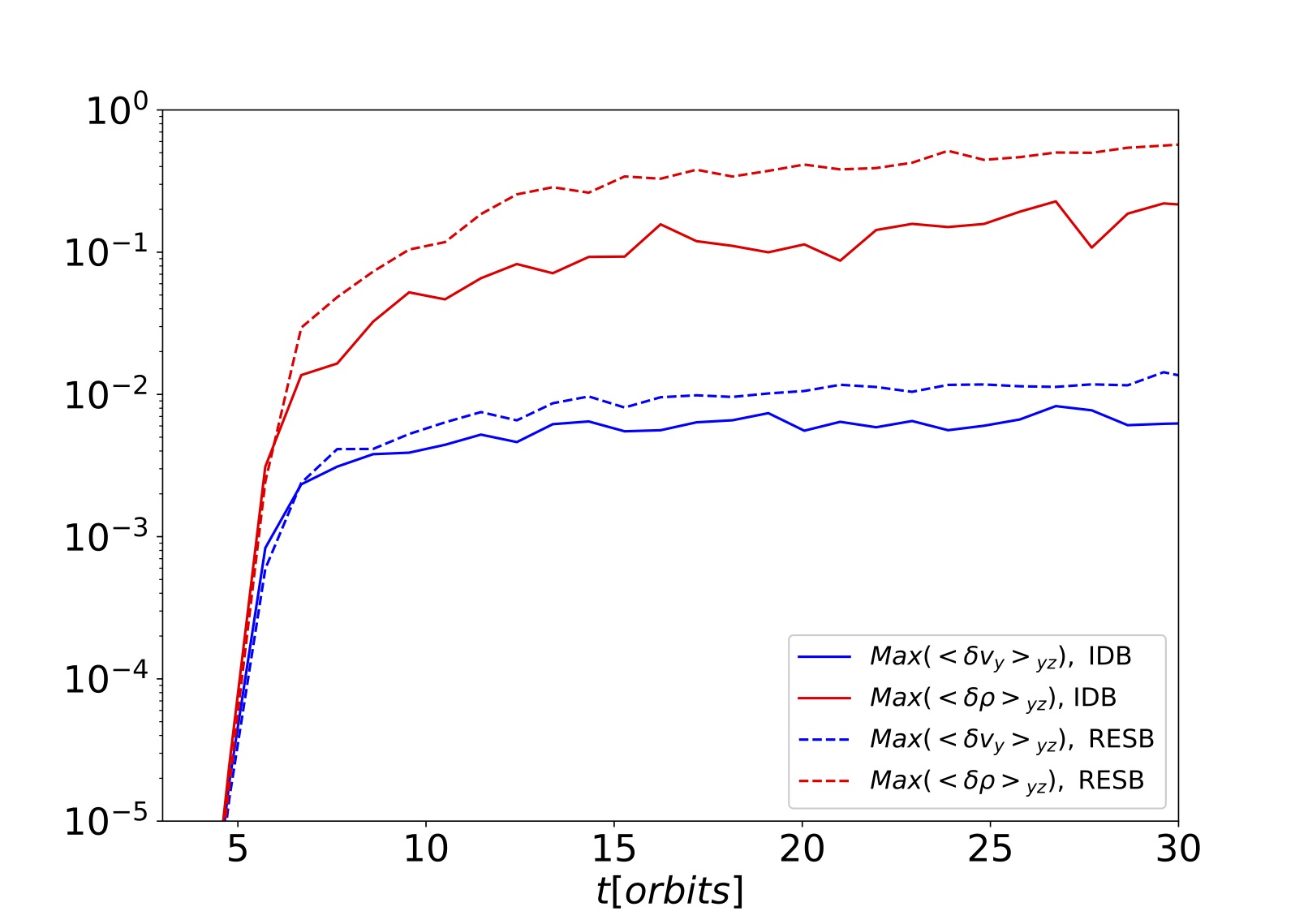}
\caption{Density and $y$-component perturbations maximum values in the $x$ direction, as a function of time, normalized with Max($v_0$) and Max($\rho_0$) respectively. The dashed lines correspond to simulation IDB while the solid lines correspond to simulation RESB. }
\label{vphi_rho_max}
\end{figure}

\begin{figure}
\hspace{-0.3cm}
 \includegraphics[width=87mm]{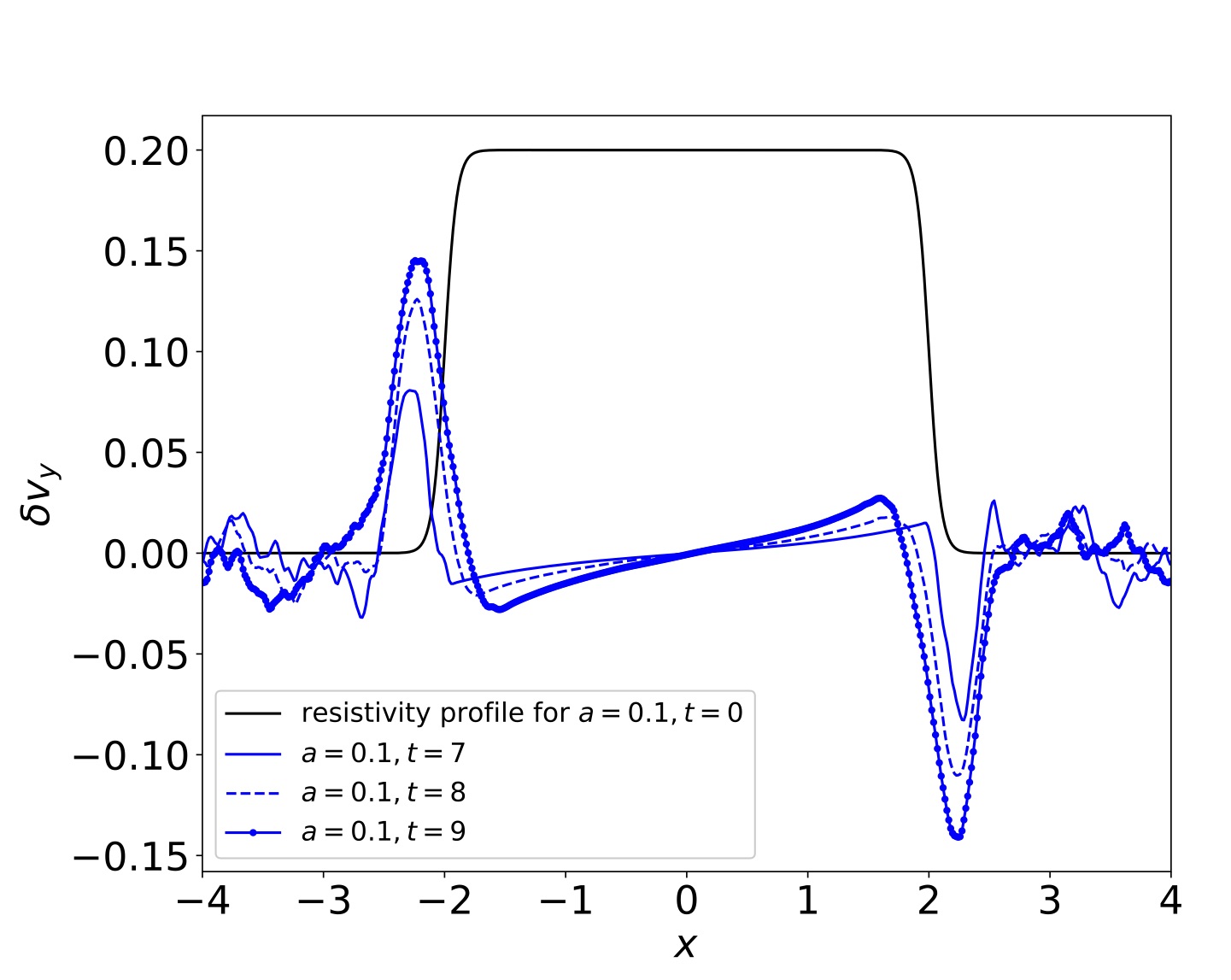}
\caption{Resistivity and $\delta v_y=v_y-v_0$ as a function of the x coordinate for simulation RESB, when MRI is kicking in and the velocity fluctuations form:  solid (t=7 orbital times)), dashed (t=8 orbital times), solid and dotted (t=9 orbital times) marks. The figure shows the peaks form in correspondence of the resistivity gradient.}
 \label{res_and_vel_prof}
\end{figure}

The timescale for the formation of density and stream structure is very similar and in order to prove they are strictly connected with the MRI linear development (before saturation), we plot $\delta  \rho_{max}(t)=(\rho(t)-\rho_0)_{max}/\rho_0$ and $\delta v_y(t)_{max}=(v_y(t)-v_0)_{max}/\mathrm{MAX}(v_0)$ maximum values in $x$ direction as a function of time, both for IDB and RESB (Fig.  \ref{vphi_rho_max}), where the subscript $0$ indicates those are the initial values. 
The logarithmic scale on the vertical axis shows the growth rate of both the velocity and density perturbation is exponential and the saturation value is just slightly larger in the resistive case, supporting the idea that the resistivity gradient alters the MRI linear evolution. 
In Fig. \ref{res_and_vel_prof} we show $\delta v_y$ profiles at different times, for simulations RESB. 
The velocity fluctuations are about $\sim30\%$ of the initial local shearing flow. The amplitude of these velocity perturbations, with respect to the background shear, is comparable to the peak velocity fluctuation amplitude in the center of the ideal part of the domain. As shown by \citet{neumanBlackman:2017}, velocity structures can appear in the turbulent domain which, for sufficiently high Reynolds and magnetic Reynolds numbers should contribute to form smaller turbulent structures.
While this is a very interesting and important topic and deserves to be investigated further, in our simulation we can clearly see the modification of the ``local Keplerian flow'' (in the shearing box the rotation is approximated with a linear velocity profile) occurs in correspondence of the resistivity gradient. 
Similar analysis on RESC shows the steeper the transition for the resistivity profile, the more localized the velocity perturbations are. 
\subsection{Comparison with similar findings in the literature.}
This density peak has been observed in other simulations before, e.g. in \citet{Katoetal:2009,Katoetal:2010,Katoetal:2012} and in \citet{Faure_et_al:2014}, even with a more realistic resistivity depending on the temperature profile. Even if the setups in these works are different, the density feature can be explained as the effect of the MRI itself developing in a setup with a non uniform resistivity \citep{Katoetal:2009, Katoetal:2010}. 
\citet{Katoetal:2010} noticed the angular velocity profile of gas is modified when MRI is excited non-uniformly in a part of a disk. By the end of the linear phase of the MRI, the velocity profile (in the shearing direction) exhibits a rigid-rotation in correnspondence of the transition between the active and inactive regions. 
Indeed we expect the MRI not to be active in the region where the macroscopic magnetic Reynolds number $R_m<1$. In particular following \citet{Fleming_Stone_Hawley:2000}, these are the region where $R_m \sim R^{cr}_m$.
In our simulation RESB, as opposite to \citet{Katoetal:2009}, the modification to the initial velocity profile seems to accelerate the shearing velocity at the transition region. This is due to the fact that the net effect of the MRI is to redistribute angular momentum within the box: the MRI inactive layer, close to regions where the MRI is active, are dragged by the nearby active layer. Notice while there is no actual angular momentum transport in a shearing box simulation, the angular momentum is redistributed by MRI over the box. If this active region is characterized by higher speed than the local dead zone, the latter is speeded up as in our case and, viceversa, the outer inactive layer is slowed down by the slower MRI active layer \citep{Katoetal:2010}.
\subsection{Density accumulation as a diffusion process.}
Another way to understand the density enhancements is through a turbulent diffusion process linked to the turbulence strength $D \propto \alpha$ (see e.g. \citet{Kalinske:1943}), where the diffusion coefficient enters into the evolution of the density as: 
\begin{equation}
\label{diffusionEQ}
d\rho/dt \sim \nabla\cdot(D\nabla \rho)
\end{equation}
In the $x$ (radial) direction, eq. \eqref{diffusionEQ} becomes
$$d\rho/dt \sim (dD/dx)(d\rho/dx) + D d^2\rho/dx^2.$$
The first term on RHS can be interpreted as an advection equation with the 
advection velocity of $-dD/dx$. In the transition region, $dD/dx$ is large because, as shown in Fig. \ref{alpha}, \,
$\alpha$ varies significantly across the transition region. As $D$ is large in the active zone but 
small in the dead zone, this can produce a net mass flux from the active zone to 
the dead zone. Once the pressure and so the density distribution is altered, the 
disk adjusts itself so that the pressure gradient is balanced by the Coriolis 
force, and the quasi-steady state is achieved.
%

\subsection{\label{subsect:CI} Spectra in the resistive models.}
\label{subsect:CI}
\begin{figure}
\includegraphics[width=90mm]{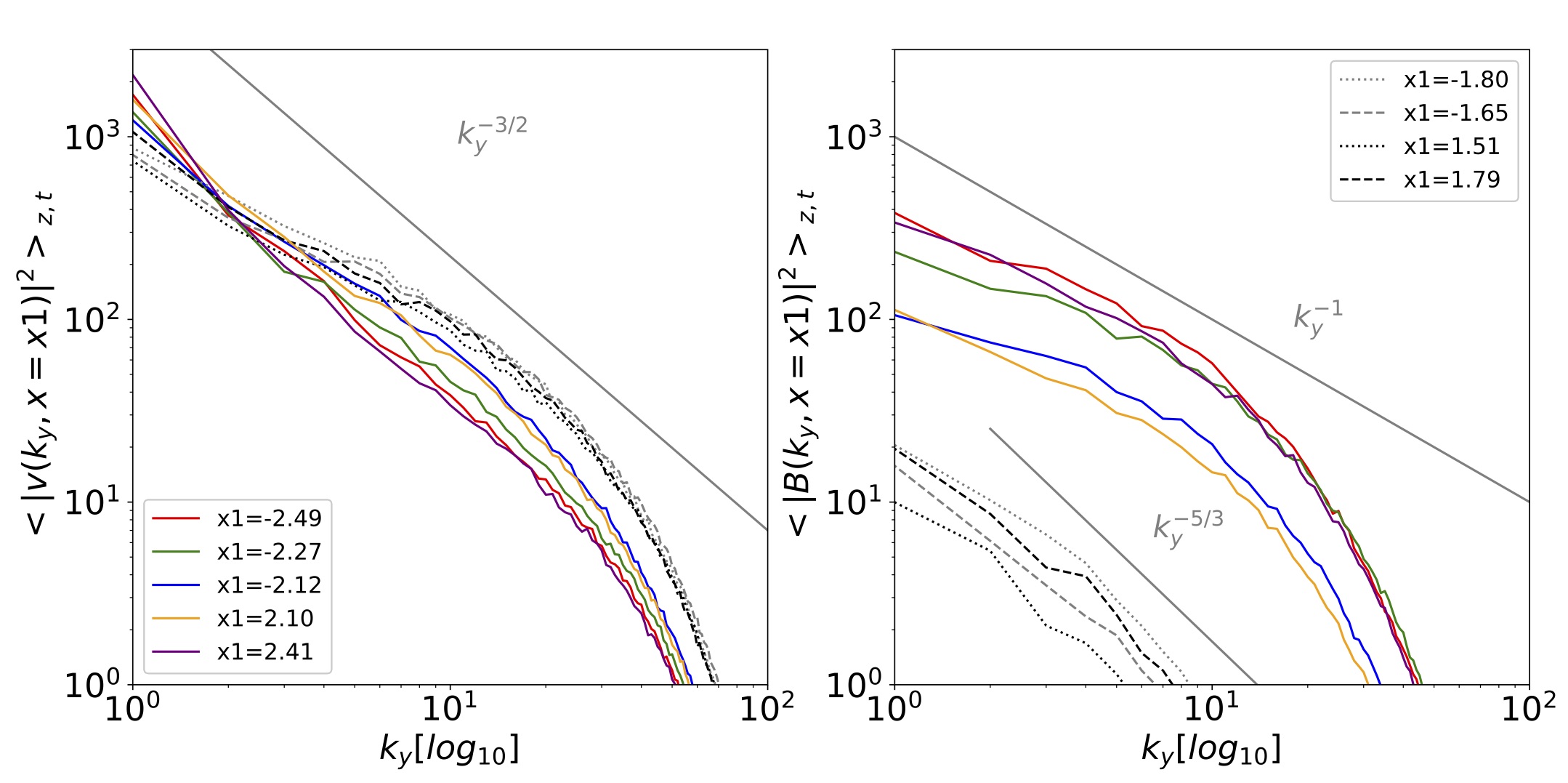}
\caption{1D spectra in the $y$ direction for simulation RESB, averaged in the vertical direction and time (last 10 orbits). Different colors label the location in the $x$ direction. Colors label the active layers, while black and grey dashed and dotted lines label the resistive layers. For a better visualization the legend is spread among the two panels and refers to both of them.}
 \label{spectraVI}
\end{figure}

We investigate the spectral features in the shearing direction, considering averages as defined previously (see Eqs. \eqref{eq:spectra}-\eqref{eq:spectra3}), i.e. fields averaged in z, and then spectra at different positions $x_1$ averaged over time ($t=25-45\ orbits$).
In Fig. \ref{spectraVI} (left), we can see the velocity spectra for RESB simulation, where colors label different values of x. i.e. specific distances from the boundaries between dead and active zones. When compared to the previous kinetic energy spectra,
we immediately notice a difference between the active and dead zones. As before, in the active regions, the spectra are compatible with a $-3/2$ slope. Velocity field fluctuations are present also in the dead zone, but the kinetic energy is strongly suppressed there at larger scales, while smaller scale fluctuations appear to propagate into the dead zone, where the spectra become flatter. This must be due to the interplay of the inhomogeneously developing MRI instability and the propagation of density fluctuations.
The RMS of the magnetic field is plotted in Fig. \ref{spectraVI} (right). Compared to the ideal simulation IDB, the magnetic energy is strongly quenched in the dead zone. In the active region the spectra are flat. As before it is difficult to really identify a power law, but the scaling like $k_y^{-1}$ for about one decade from the large injection scales is not far off. The dead layers are characterized by significantly lower magnetic energies (almost two orders of magnitude less), but there is a steeper slope, characterized by a Kolmogorov type spectrum for about one decade. This suggests that the dynamics in the dead zone is not dominated directly by energy injection from the MRI, but rather a more complex process involving injection via the velocity field and density that penetrates the dead zone.


\section{\label{sec:IV}Conclusions.}

The main goal of our study was to quantify the turbulence and dynamics in the proximity of the radial transition region in the saturated phase of the MRI. In this work we performed shearing box simulations using the Athena++ code, with a non uniform ohmic resistivity in the radial direction.Our analysis shows that the most important parameter to determine the turbulence behavior is the actual value of the resistivity in the dead zone, while the thickness of the transition region does not significantly affect the turbulence values. 
 In addition, from the computational point of view, this work confirms that resolving the transition region is not fundamental to determining the physics across the boundary itself, which supports the validity of results on the dynamics of the transition region in global domains, where small scales are not actually resolved.
Global simulations are extremely expensive, being devoted to capture matter accretion and the effect of winds, the latter being particularly relevant for the transport in the turbulent inhibited regions. Resolution required to study the physics and the dynamics at this key region, especially for different plasma parameters, can only be achieved in local simulations. 

\subsection{Findings and results.}
For comparison we performed ''ideal'' simulations,  for which for our fiducial model is IDB. In these runs an actual resistivity is provided by a finite spatial grid. We observe  a saturation of the viscous stress $\alpha \sim 0.035$.
Our resistive models are characterized by a non-uniform explicit ohmic resitivity in the radial direction. The radial profile of the resistivity transitions linearly from zero to $\eta$, defining two active zones, and a dead zone in the center of the simulation domain. Our fiducial model RESB shows a saturation phase (lasting about 30 orbits), during which the magnetic contribution is the most relevant for the Stress tensor. The magnetic field dissipates quickly in the dead zone, but the velocity perturbation propagates beyond the transition region into the resistive area, sustaining turbulence with an effective $\alpha$ at the center of the resistive region of  $\sim 20 \%$ of the (average) ideal MHD region, both for run RESB as well as for run RESC (the latter is characterized by a shorter, not resolved transition region for comparison). In simulation RESA though, for which the explicit resistivity value $\eta$ is 10 times larger than in RESB, the dead zone region is magnetically dead. For all the resistive simulations, the turbulence level in the active areas of the domain rises to $\sim 30 \%$ of the ideal MHD region, up to one scale height deep into the resistive region (within $|x|=1.5$ and $|x|=2.5$ in Fig. \ref{alpha}), depending on the explicit resistivity value.
We observe a sharp transition in the Maxwell tensor from the MRI turbulent active areas to the dead zone. On the other hand, the sharper transition does not affect significantly the effective viscosity. 
\newline
\newline
The one dimensional Fourier spectra in the shearing direction, can be fitted with a slope compatible with a $-3/2$ scaling; the magnetic field exhibits a flat spectrum at large scales, then falls off at values consistent with the estimated numerical dissipation scales. Energies at large scales are comparable for the velocity and magnetic field.
In our resistive fiducial model (RESB) the velocity spectrum can be again fitted with a $k_y^{-3/2}$ slope in the active layers, while it flattens at large scales for the dead layers, resembling a  $k_y^{-1}$ slope. While in the active region the spectra are flat, scaling like $k_y^{-1}$ for about one decade from the large injection scales, the dead layers are characterized by significantly lower energies, almost two of magnitude less than the ideal regions. In conclusion, the spectra do not reveal significant features in transition between active, and the non active zone, while the magnetic energy is clearly quenched at all scales in the resistive region and in the transition region.
\newline
\newline 
\subsection{Comparison with the literature.}
In the ''ideal'' simulations we observe a saturation of the viscous stresses comparable with and in agreement with previous literature (see e.g. \citet{Hawley:1995}), where a similar resolution is adopted in the vertical direction (64/H). The main contribution to the transport comes from the Maxwell tensor, as expected. 
We observe that the result on the ''radial'' transition to the dead zone, that retain some turbulent activity and fluctuations of the Reynolds stress is similar to what found in \citet{Fleming_Stone:2003} for the vertically stratified shearing box, i.e. the vertical variation of the ionization degree while, in our case, the non uniform resistivity takes into account the variation of the ionization degree in the radial direction.
\newline
\newline
Global dynamics can change the thermal structure of  the disks. For example, \citet{Faure_et_al:2014}, using a model where the dissipation in the system has a feedback on the temperature profile and a simple prescription for radiative cooling, pointed out the importance of heating caused by waves propagating adiabatically through the dead zone, and dissipating as weak shocks. This result is important when the heating is taken into account (so for more realistic models): changing the temperature, these waves can also change the resistivity profile, and so the location of the transition region. These global effects will be studied in our future papers.
\newline
\newline
Consistently with \citet{Katoetal:2009}, our resistive setups show the resistivity gradient alters the angular momentum redistribution at the boundaries between the active and the dead zone. In the stationary phase the strong velocity gradient can be express in terms of the resistivity gradient, which is significantly high in the transition region. While this is not the primary explanation for density accumulation and planetesimal growth at the boundary of the dead/active zone, this might be a competing  important effect.

\subsection{Relevance of this work and limitations}

As found by \citet{neumanBlackman:2017} the critical Reynolds numbers for which turbulence can be sustained in the active zones, is still a matter of debate. High numerical resistivity might induce in wrong consideration of the parameters determining the sustainability of MRI and its saturation values.
\newline
\newline
This work is relevant in the context of protoplanetary disks, for which the role of non-ideal MHD effects has been recognized in the region $r >1AU$, where dust grains are most probably trapped and evolve into planetesimals. 
An example is the work by \citet{Okuzumi:2013} which presented simple scaling relations for the planetesimal stirring rate in turbulence driven MRI, taking into account the stabilization effect of an ohmic resistivity. These findings motivate an investigation of the chemistry and radiation processes required to provide the correct non ideal coefficients that are particularly relevant in affecting the turbulence dynamics, see e.g. the recent paper \citet{Gresseletal:2020}.
\newline
\newline
The combined effect radial transition/vertical stratification should also be taken into account to understand the transport and accretion in a realistic protoplanetary disk model. In particular, vertical stratification may allow the formation of the so called \textit{zonal flows} \citep{Johansen:2009, Kunz:2013, Bai_and_Stone:2014}, contributing to create denser regions at different heights, balancing the momentum equation in the disk.  As discussed in the introduction, the equatorial plane of the disk is also interested by other non ideal effects than ohmic resistivity, and connected with the chemical and radiation processes occurring within the disk and in the central star (see e.g. \citet{Okuzumi_Hirose:2011,Gressel_et_al:2015,Xu_Bai:2016}).

\acknowledgements
We would like to thank Prof. Kazunari Iwasaki for discussions and insights on accretion in astrophysical disks and simulations. FP would like to thank Prof. Marco Velli and Dr. Silvio Cerri for discussions on MHD turbulence and energy transfer. FP would also like to thank Dr. Neal Turner for illuminating discussions on the results of this manuscript.
The simulations presented in this paper were performed on Perseus supercomputer in Princeton University (https://researchcomputing.princeton.edu/systems-and-services/available-systems/perseus).
This research was supported in part by the National Science Foundation under Grant No. NSF PHY-1748958.
KT was supported by Japan Society for the Promotion of Science (JSPS) KAKENHI Grant Numbers 16H05998, 16K13786, 17KK0091, 18H05440. KT also acknowledges support by MEXT as ''Program for Promoting Researches on the Supercomputer Fugaku'' (Toward a unified view of the universe: from large scale structures to planets).
\bibliographystyle{apj}
\bibliography{Bib}

\end{document}